\documentstyle[prl,aps,floats,epsf,multicol]{revtex}
\tighten

\renewcommand{\refname}{}
\newcommand{\biblabel}[1]{[#1]} %
\renewcommand{\references}{%
\ifpreprintsty
\vspace*{-0.1 truein}
\hbox to\hsize{\hss\large \refname\hss}%
\else
\vskip3pt
\hrule width\hsize\relax
\vskip -0.2in
\fi
\list{\biblabel{\arabic{enumiv}}}%
{\labelwidth\WidestRefLabelThusFar  \labelsep4pt %
\leftmargin\labelwidth %
\advance\leftmargin\labelsep %
\ifdim\baselinestretch pt>1 pt %
\parsep  4pt\relax %
\else %
\parsep  0pt\relax %
\fi
\itemsep\parsep %
\usecounter{enumiv}%
\def\theenumiv{\arabic{enumiv}}%
}%
\let\newblock\relax %
\sloppy\clubpenalty4000\widowpenalty4000
\sfcode`\.=1000\relax
\ifpreprintsty\else\small\fi
}

\newcommand{\smeq}{\! = \!}

\def\lsim{\hbox{\lower .8ex\hbox{$\, \buildrel < \over \sim\,$}}}
\def\gsim{\hbox{\lower .8ex\hbox{$\, \buildrel > \over \sim\,$}}}

\newcommand{\Vtf}{V_{\rm TF}}
\newcommand{\hV}{\hat \Vtf}
\newcommand{\Vzr}{V_{\mbox{zero-range}}}

\newcommand{\bq}{{\bf q}}

\newcommand{\br}{{\bf r}}

\newcommand{\Vext}{V_{\rm ext}}
\newcommand{\Veff}{V_{\rm eff}}
\newcommand{\Vsc}{V_{\rm sc}}

\newcommand{\ETF}{E_{\rm GTF}}
\newcommand{\FTF} {{\cal F}_{\rm GTF}}
\newcommand{\TTF}{{\cal T}_{\rm TF}}

\newcommand{\FLDA}{{\cal F}_{\rm DFT}}

\newcommand{\ELDA}{E_{\rm DFT}}

\newcommand{\Eext}{{\cal E}_{\rm ext}}
\newcommand{\Ecoul}{{\cal E}_{\rm coul}}

\newcommand{\Exc}{{\cal E}_{\rm xc}}
\newcommand{\Eop}{{\cal E}_{\rm 1p}}

\newcommand{\Etot}{{\cal E}_{\rm tot}}
\newcommand{\nLDA}{n_{\rm DFT}}

\newcommand{\nTF}{n_{\rm GTF}}
\newcommand{\nosc}{{\tilde n^{\rm osc}}}

\newcommand{\e}{\epsilon}
\newcommand{\ve}{\varepsilon}

\newcommand{\kf}{k_{\scriptscriptstyle F}}
\newcommand{\Ef}{E_F}
\newcommand{\Eth}{E_{\rm Th}}
\newcommand{\bu}{{\bf u}}
\newcommand{\bk}{{\bf k}}

\epsfclipon

\begin{document}
 

\draft

\title{Interactions in Chaotic Nanoparticles: \\
       Fluctuations in Coulomb Blockade Peak Spacings}

\author{Denis Ullmo$^{1}$ and Harold U. Baranger$^{2}$} 

\address{
$^{1}$Laboratoire~de~Physique~Th\'eorique~et~Mod\`eles~Statistiques~(LPTMS),~%
91405~Orsay~Cedex,~France}

\address{$^{2}$Department~of~Physics,~Duke~University,~Box~90305,~%
Durham~NC~27708-0305,~USA}

\date{March 1, 2001}

\maketitle

      \begin{abstract}
We use random matrix models to investigate the ground state energy of
electrons confined to a nanoparticle.  Our expression for the energy
includes the charging effect, the single-particle energies, and the
residual screened interactions treated in Hartree-Fock.  This model is
applicable to chaotic quantum dots or nanoparticles---in these systems the
single-particle statistics follows random matrix theory at energy scales
less than the Thouless energy. We find the distribution of Coulomb blockade
peak spacings first for a large dot in which the residual interactions can
be taken constant: the spacing fluctuations are of order the mean level
separation $\Delta$.  Corrections to this limit are studied using the small
parameter $1/\kf L$: both the residual interactions and the effect of the
changing confinement on the single-particle levels produce fluctuations
of order $\Delta/\sqrt{\kf L}$.  The distributions we find are
significantly more like the experimental results than the simple constant
interaction model.
\end{abstract}

\pacs{PACS numbers: 73.23.Hk, 05.45.Mt, 71.10.Ca, 73.20.Dx}


\narrowtext

\widetext
\vspace*{-0.1in}
\begin{multicols}{2}

\section{Introduction}
   \label{sec:intro}

The ground state properties of electrons in nanoparticles or
quantum dots---finite systems of fermions interacting via Coulomb
forces---remain incompletely understood. One valuable way to probe
these systems is via electrical transport (for reviews see Refs.
\onlinecite{GrabDev92,MesoTran97,FerryGood97,Timp99}). The dominant effect
is the suppression of the conductance $G$ because of the charging energy needed
to add an electron to the nanoparticle and so allow current to flow. This
{\em Coulomb blockade (CB)} is an essentially classical effect, and the
charging energy is simply $E_C \smeq e^2/2C$ where $C$ is the capacitance
of the nanoparticle. The blockade can be lifted by adjusting the potential
on the particle---with a gate voltage $V_g$, for instance---so that the
energy for $N$ electrons is the same as that for $N \!+\! 1$. The
conductance through the nanoparticle has peaks at these degeneracy points;
in fact, $G(V_g)$ is periodic with a spacing of $e^2/C$ between the peaks
\cite{GrabDev92,MesoTran97}.

At low temperature the electron remains coherent in the dot and
so quantum interference effects must be included in the Coulomb
blockade \cite{KastnerRMP92,KouwenetalRev97,RalphRev97,AleinBrouwGlaz01}. Two
main effects occur: (1)~the peak {\em heights} fluctuate from
peak to peak because of varying coupling between the wavefunction
in the nanoparticle and the leads, and (2)~the {\em spacings}
between the peaks fluctuate because of interference contributions to
the ground state energy. Peak heights have been studied first theoretically
\cite{JalStoneAlh92,Stopa96,Stopa98,Hackenbroich97,Alhassid98,Vallejos98,Vallejos99,E2NBar99,Kaplan00,E2NBar01,MarcusPatel98a}
and then experimentally \cite{MarcusPatel98a,ChangPeaks96,MarcusFolk96};
agreement is good once effects of temperature
\cite{Alhassid98,MarcusPatel98a} and classical dynamics
\cite{E2NBar99,Kaplan00,E2NBar01} are taken into account.

In this paper we focus on quantum effects in the peak spacings. These
are directly related to the interference contributions to the ground state
energy. One obvious contribution is the single-particle energy levels caused
by the confinement in the quantum dot. But there may also be contributions
coming from the interactions in the dot. We will show that for typical
parameters characterizing experimental quantum dots or nanoparticles the
contributions coming from the interactions must be included.

The simplest treatment of Coulomb blockade peak spacing fluctuations
results from writing the ground state energy as the sum of the classical
electrostatic energy and the energies of the occupied single particle states
\cite{KastnerRMP92}.
Such an expression for the ground state energy of a many Fermion system
has a long and venerable history in atomic and nuclear 
physics~\cite{Mayer,Latter,Bohigas76}.
Using $\cal{E}$ for the energy in the presence of the gate voltage while
reserving $E$ for $V_g \smeq 0$, we have for $N$ particles
   \begin{eqnarray}\label{eq:CIenergy}
    {\mathcal E}_N (V_g) & = & E_N - (C_g/C) eNV_g \\
       E_N & = & (e^2/2C) N^2 + \sum_{{\rm occupied}\ i\sigma}
				 \epsilon_{i\sigma}
       \nonumber
   \end{eqnarray}
where $C_g$ is the capacitance of the gate to the dot and 
$\{\epsilon_{i\sigma}\}$
are the single-particle energy levels.	This treatment, known as
the constant-interaction model (CI), is still an important point of
reference in the field.  The gate voltage $V_g^*$ at which a peak in
the conductance occurs is given by the condition ${\mathcal E}_N (V_g^*)
\smeq {\mathcal E}_{N+1} (V_g^*)$.  The spacing between two peaks is, then,
proportional to the (discrete) inverse compressibility $\Lambda$
   \begin{eqnarray}\label{eq:sdef}
      \lefteqn{
        (V_g^*)_{N \rightarrow N+1} - (V_g^*)_{N-1 \rightarrow N} \propto }
      \\
       & & \qquad \qquad \qquad \qquad
             E_{N+1} + E_{N-1} - 2 E_N \equiv \Lambda_N \;.
          \nonumber
   \end{eqnarray}
In the case of the CI model, one gets the simple prediction \cite{Sivan96}
   \begin{eqnarray}\label{eq:CIprediction}
    \Lambda_N = {e^2}/{C} \qquad \qquad \qquad \qquad \qquad
	&& \mbox{for odd $N$} \nonumber \\ \Lambda_N = {e^2}/{C} +
     (\epsilon_{N/2+1}-\epsilon_{N/2}) \qquad &&
      \mbox{for even $N$} \; .
   \end{eqnarray}
Note the drastic odd/even effect in the quantum correction to the classical
result $e^2/C$ because of the spin of the electron.

A number of experiments have studied the distribution
of CB peak spacings in semiconductor quantum dots
\cite{Sivan96,Simmel97,MarcusPatel98,Simmel99,LuschEnsslin01}, a
particularly flexible and convenient type of nanoparticle.  The overall
result has been striking {\it disagreement} with the CI model.  The main
discrepancy is that there is no sign of the strong odd/even effect
predicted by Eq.~(\ref{eq:CIprediction}), though there is evidence of
a weak odd/even effect \cite{LuschEnsslin01,BarOng01}.  In addition,
in some of the experiments the widths of the observed distributions are
significantly larger than what is predicted using the CI model and assuming
the single-particle energy levels $\epsilon_i$ are distributed according
to the appropriate Gaussian random matrix ensemble.  These discrepancies
indicate unambiguously that interaction effects play a significant role
in the fluctuating part of the Coulomb Blockade peak position.

Theoretically accounting for this disagreement is difficult because of two
intrinsic characteristics of the problem: (1)~the fairly large number of
electrons contained in the dot (typically $N \!\sim\!  100$-$1000$), and
(2)~the long range character of the Coulomb interaction which requires
that screening has to be correctly taken into account.  The first of these
points makes it delicate to extend to realistic dots the results obtained
by exact multiparticle diagonalization \cite{Sivan96,Berkovits98}, which
can only treat problems with a few electrons ($\lsim 15$).  To reach the
experimental number of particles in a numerical approach, one has to rely on
some self-consistent technique. One approach is Hartree-Fock which has been
recently used in quantum dots \cite{Walker99a,Walker99b,Cohen99,Ahn99}.
Then, however, the second point comes into play.  Indeed, screening
is a higher-order interaction effect, and it is well known for the
Coulomb interaction that the exchange terms can be significantly
over evaluated in self-consistent Hartree-Fock (for the bulk see
Ref. \onlinecite{AshcroftMermin1} but this applies also for restricted
geometries).  This makes it rather difficult to interpret numerical
self-consistent Hartree-Fock calculations using the bare Coulomb interaction
as input.  Another possible self-consistent approach is, of course, density
functional theory in the local density or local spin density approximation
\cite{ReimannMann97,MartinLeb98,Wing99}.  Here the screening is properly
handled but approximations are made in the exchange term.  Since, as we
explain in detail in this paper, the subtle interplay between exchange,
residual direct interactions, and single-particle level fluctuations is
critical in describing quantum dot physics, results from this method are
also difficult to generalize.

An alternative theoretical approach is to formally carry out
an RPA screening calculation and then use a short-range screened
interaction in the subsequent evaluation of the magnitude of various
effects~\cite{Prus96,Berkovits97,BlantMirMuz97,AleinBrouwGlaz01}.  
The nature of the
screened interaction used is, then, important.  Since experimentally the
system size is much larger than the screening length, the most natural choice
for the interaction is the bulk RPA screened interaction. For momentum
smaller than $2 \kf$ this is correctly approximated by the Thomas-Fermi
expression. Throughout this paper we consider two dimensional systems,
in which case
   \begin{eqnarray}
  \Vtf(\br) & = & \int \frac{d\bq}{(2\pi)^2} \ \hV (\bq) \
      \exp[i\bq\!\cdot\!\br] \ , \nonumber \\
  \hV (\bq) & = &
    \frac{\nu^{-1}}{1+ r_s^{-1} (|\bq| /\kf)/\sqrt{2} } \
        \label{eq:Vtf}
   \end{eqnarray}
where $\nu$ is the mean density of states (including spin degeneracy),
and $r_s \smeq r_0 / a_0$ with $\pi r_0^2$ the average area per electron
and $a_0$ the Bohr radius in the semiconductor.  It should be born in
mind that although $\hV$ is proportional to $r_s$ for small $r_s$, it
becomes largely independent of $r_s$ when $r_s \gsim 1$, which is the
case experimentally. In fact, within any RPA-like approach the interaction
cannot become stronger than
   \begin{equation} \label{eq:local_int}
      \Vzr(\br) = \nu^{-1}
          \delta(\br)
   \end{equation}
which we refer to as the {\em zero-range} limit of the potential.  Once screening
is included, the problem is that the interaction is small, in a sense
that we shall make more explicit below, even in the strongest zero-range limit.  
Thus the self-consistent approach should give roughly
the same answer as the CI model with first-order perturbation in the
screened interaction added.  This is exactly the picture one has of weakly
interacting Landau quasi-particles, with the obvious drawback that it is
not straightforward to extract large contributions using this approach
(see e.g. Refs. \onlinecite{Prus96,Berkovits97,BlantMirMuz97}).

In order to obtain a large interaction effect, one possibility is to give
up the Fermi-liquid-like approach and to assume, for instance, that for some
reason screening is not as efficient in a nanoparticle as in the bulk, and so
conclude that large zero-range interactions, not amenable to a perturbative
treatment, should be used~\cite{Sivan96,Berkovits98,Prus96,Berkovits97}.
Before taking this drastic step, it seems reasonable, however, to see how
far one can go within the more standard Fermi liquid description.  Indeed,
even in this framework, we know that the CI model is inadequate.

An accurate treatment of the spacing or inverse compressibility certainly
requires that the average effect of the residual interactions be added to
the CI model. By ``residual interactions'' we mean interactions beyond the
simple classical effect taken into account in the constant interaction
model and, in particular, interactions between the quasiparticles
after the bare electrons are dressed through screening.  While they
make a smaller contribution to the total energy than those considered
in Eq.~(\ref{eq:CIenergy}), residual interactions make a contribution
of order the mean level separation to the second difference $\Lambda$,
comparable to the single-particle contribution in Eq.~(\ref{eq:CIprediction})
\cite{Prus96,BlantMirMuz97,BrouwOregHalp99,BarUllGlaz00,MesoStoner00,JacquStPRL00}.
In the limit of a large nanoparticle---one whose typical dimension
$L$ is many times the electron wavelength, $\kf L \gg 1$---only the
average effect of residual interactions needs to be added to the CI
model \cite{BlantMirMuz97,BrouwOregHalp99,BarUllGlaz00,MesoStoner00}.
We shall below refer to this approximation as the {\em constant exchange}
model.  The intuitive argument for this is that the effect of residual
interactions involves integrating wavefunctions over the entire volume of
the nanoparticle and so self-averages: the particular characteristics of
any given wavefunction are simply integrated out.  Even if one neglects the
fluctuations of the residual interaction terms, because of fluctuations of
the single-particle levels, the ``constant exchange'' contribution can modify
the total spin of the nanoparticle by favoring non-trivial occupation of the
single-particle orbitals and, as a consequence, will significantly modify the
fluctuations of the peak spacings~\cite{Prus96,BlantMirMuz97}.  The ground
state spin of a large nanoparticle in this regime has been examined in
detail \cite{BrouwOregHalp99,MesoStoner00,JacquStPRL00,JacquStPRB01}, but
curiously the corresponding distribution of Coulomb blockade peak spacings
has not appeared explicitly (see however Ref. \onlinecite{BlantMirMuz97}).
Therefore, before turning to effects caused by fluctuations in the
interactions---our main interest in this paper---we present below the
distribution in the constant exchange limit ($\kf L\gg 1$) to use as a
reference point.

In the experiments, however, $\kf L$ is not very large because of the
need to have the mean level separation larger than the temperature.
This suggests that the variations in interactions caused by properties of
individual wavefunctions are important.  The main goal of this paper is,
therefore, to study the effect of fluctuations in the electron-electron
interactions on the distribution of CB peak spacings.

We limit our study here to the case of {\em chaotic dynamics} within the
dot, for which a random matrix description of the non-interacting limit
can be used \cite{GutzBook}. Much as in the case of diffusive transport,
completely chaotic dynamics introduces an important additional energy
scale (and only one): the Thouless energy, $\Eth \smeq \hbar /\tau$, is
the inverse of the transit time across the dot. It is generally larger
than the mean level separation $\Delta$ but smaller than the Fermi energy;
we will assume $\Delta \!\ll\! \Eth \!\ll\! \Ef$. On energy scales less
than $\Eth$, a universal random matrix description holds while for larger
energies (shorter times) the individual dynamics of the system comes into
play. We will see that taking account of this additional energy scale is
critical in describing CB peak spacings in nanoparticles.

We focus on two effects in particular: (1)~the fluctuation in the residual
Hartree and Fock contributions to the ground state energy and (2)~the
change in the single-particle energies because of changes in the mean
field potential as electrons are added to the dot \cite{BlantMirMuz97},
an effect we refer to as ``scrambling''.  For two dimensional quantum dots,
the changes in energy are evaluated to leading order in the smallness of the
dot, the constant exchange model being used as the starting and
reference point.  We find that both effects contribute terms which are of
order $\Delta /\sqrt{\kf L}$, up to logarithmic corrections.  The resulting
distribution of CB peak spacings looks more like the experiment---it has
a more Gaussian and symmetric form, for instance---but still deviates from
the measurements in significant ways.

The paper is organized as follows.  Our starting point is a
semiclassical expansion of the ground state energy described in Section
\ref{sec:approach}---an expansion in $1/\kf L$. The important large dot
limit is presented in Section \ref{sec:largedot}.  The next two sections
discuss the two main issues---fluctuations of the residual
interaction contributions (Section \ref{sec:residualint}) and scrambling
(Section \ref{sec:scrambling}). Section \ref{sec:level2} takes a first
step beyond the Gaussian model introduced in Section \ref{sec:residualint}.
Finally, our conclusions, including a discussion of the experimental results,
appear in Section \ref{sec:conclu}.

\section{Approach: Semiclassical Corrections}
   \label{sec:approach}

In the Coulomb blockade through quantum dots we are faced with a classical
theory which works remarkably well to which we want to add quantum effects
to leading order. The small parameter is the standard semiclassical one,
$\hbar$. In our context, this is equivalent to an expansion in $1/\kf L$
where $\kf$ is the Fermi wavevector and $L$ is the typical size of the
nanoparticle. Another useful way to view the corrections is in terms of
the dimensionless conductance, $g$, given as the ratio of the transit
rate $\hbar v_F/L$ to the mean level separation $\Delta$; in this case
the expansion is in terms of $1/g$, as in the diffusive mesoscopic
regime \cite{AltSimHouches}.  For 2-dimensional systems, such as the ones 
we consider here, these two parameters are proportional ($g=\kf L/2\pi$).

We proceed by using the method of
V. M. Strutinsky~\cite{Strutinsky68,Strutinsky72} in which the dependence
of many-body ground state quantities on particle number can be decomposed
into an average and a fluctuating part.  While the average part varies
smoothly with particle number, the fluctuating part reflects the shell
structure of the system.  The smooth part is the bulk energy per unit
volume integrated over the finite-size system, and the oscillating
contributions come from quantum interference effects explicitly
caused by the confinement.  By supposing that the smooth part is known
while the unknown oscillatory contribution is a correction, Strutinsky
introduced a physically motivated systematic approach to obtain the shell
corrections~\cite{Strutinsky68,Strutinsky72}.  This shell correction method
is essentially a {\it semiclassical} approximation. It rests on the fact
that the number of particles in the system considered is large, rather than
on the interaction between the particles being weak. (One must, of course,
work in a regime where the smooth starting point is basically valid.)
Since the quantum dots in which we are interested contain on the order of
100 electrons, they are a perfect place to apply the Strutinsky method.
We use the formulation recently developed specifically for quantum dots
in Ref. \onlinecite{UllTatTomBar01}.

Density functional theory guarantees that the ground state energy of a
nanoparticle can be written as a functional of its density, $\ELDA = \FLDA
[\nLDA ]$. Neglecting quantum interference effects in this functional
corresponds to a generalized Thomas-Fermi approximation (generalized
because bulk local exchange and correlation can be included) which can also
be written as a functional of a (smooth) density, $\ETF = \FTF [\nTF ]$.
The quantum interference part of the energy, $\ELDA - \ETF$, can be obtained
approximately by expanding $\nLDA$ about $\nTF$ and solving the resulting
density functional equations order by order in the oscillating part of the
density $\nLDA - \nTF$. To carry out this procedure an explicit form for
the exchange-correlation functional is required. Based on the local density
approximation, the result derived in Ref. \onlinecite{UllTatTomBar01} is
\begin{equation}\label{eq:EDFT}
     \ELDA \simeq \ETF +\Delta E^{(1)}+\Delta E^{(2)}
\end{equation}
where the first and second order correction terms are
\begin{eqnarray}
     \Delta E^{(1)} &=& 
      \Eop^{\rm osc} \bigl[\Veff [\nTF ]\bigr] \equiv
       \sum_{i=1}^N \delta \e_i
        \label{eq:deltaE1} \\
     \Delta E^{(2)} &=& 
      \frac{1}{2} \int d\br d\br' \,\nosc (\br)
         \,\Vsc(\br, \br') \,\nosc (\br') \; .
          \label{eq:deltaE2}
\end{eqnarray}
The first-order correction is the oscillatory part of the single-particle
energy, $\Eop^{\rm osc}$, calculated in the smooth Thomas-Fermi potential,
$\Veff [\nTF ]$; that is, it is the sum of the deviations $\delta
\e_i$ of the single-particle levels $\e_i$ from their mean values.
In the second-order correction, $\nosc$ is the deviation of the quantum
mechanical density calculated in the Thomas-Fermi potential from the smooth
starting point, $\nosc \equiv n\bigl[\Veff [\nTF ]\bigr]({\bf r}) - \nTF$,
and $\Vsc(\br, \br')$ is the screened interaction in the nanoparticle.
To the order that we are working, using $\nosc$ here is equivalent to
using $\nLDA - \nTF$ \cite{UllTatTomBar01}.

In this approach, the ground state energy is, then, the sum of a classical
contribution---the generalized-Thomas-Fermi result $\ETF $---and two quantum
contributions---$\Delta E^{(1)}$ and $\Delta E^{(2)}$.  With the inclusion
of only the first-order contribution, Eq.~(\ref{eq:EDFT}) is simply the
CI model for the ground state energy.  Because we have the corrections to
this model, we can see exactly when it is applicable.  The second-order
correction, Eq.~(\ref{eq:deltaE2}), has a natural interpretation: the
ripples in the density caused by interference (the Friedel oscillations in
this setting) interact with each other via an interaction screened by the
smooth Thomas-Fermi fluid already present.  In fact, the bulk Thomas-Fermi
interaction Eq.~(\ref{eq:Vtf}) can often be used for $\Vsc$.  Taking the
$r_s \rightarrow \infty$ limit formally yields the zero-range interaction limit
(\ref{eq:local_int}). This is equivalent to using a Hubbard Hamiltonian
on a square lattice with hopping term $t$ and interaction $U$ with $U/t=\pi$.

The result Eq.~(\ref{eq:EDFT}) is, however, inadequate for our purposes
here because of the local density approximation used.  In particular,
the spin degree of freedom is handled poorly in this approximation:
while Eq.~(\ref{eq:deltaE2}) gives the ``direct'' interaction between the
density ripples, an analogous ``exchange'' term is missed.  A better
treatment of the interactions in density functional theory could
presumably yield this additional term, by including off-diagonal terms
in the density matrix, for instance.  In fact, an RPA treatment of the
interaction in the diffusive case yields an energy with exactly this
structure~\cite{BlantMirMuz97,AleinBrouwGlaz01}.  Motivated by these
physical considerations, we simply add the exchange term by hand to the
expression for the energy $\ELDA$.

For the purposes of deriving CB peak spacing distributions, we thus
consider 3 contributions to the ground state energy: (1)~the Thomas-Fermi
energy, (2)~the deviation of the single-particle energy from its mean, and
(3)~a Hartree-Fock like treatment of the residual screened interaction.
Without the residual interactions, a $N$-particle state is, of course,
a Slater determinant in the basis of single-particle eigenstates
$\psi_i (\br )$; it is characterized by the occupation numbers $(n_{0
\sigma},n_{1\sigma}, \ldots)$ where $n_{i\sigma} = 0$ or $1$, $\sigma =
+1$ or $-1$ labels the spin degree of freedom, and $\sum n_{i\sigma} = N$.
In this occupation number representation, the expression for the energy is,
then, 
\begin{eqnarray} \label{eq:total_E}
      \lefteqn{
        \Etot(\{n_{i\sigma}\}) = \ETF
         + \sum_{i \sigma} n_{i\sigma} \delta \epsilon_i } \\
       && \qquad \qquad
         + \frac{1}{2} \sum_{i\sigma,j\sigma'} n_{i\sigma} M_{ij}
         n_{j\sigma'}
           - \frac{1}{2} \sum_{i,j;\sigma} n_{i\sigma} N_{ij} n_{j\sigma}
       \nonumber
\end{eqnarray} 
where 
\begin{eqnarray}
      M_{ij} & \equiv & \int d\br d\br' \left| \psi_i(\br) \right|^2
         \Vtf(\br-\br') \left| \psi_{j}(\br') \right|^2
      \label{eq:defM} \\
       N_{ij} & \equiv & \int d\br d\br' \psi_i(\br) \psi_{j}^*(\br)
         \Vtf(\br-\br') \psi_{j}(\br') \psi_{i}^*(\br')
      \label{eq:defN}
\end{eqnarray} 
are the direct and exchange contributions, respectively.  This is the
starting point for our study of CB peak spacings.

In order to derive the statistical properties of the CB peak spacings
from Eq.~(\ref{eq:total_E}), the statistical properties of the
single-particle eigenvalues and eigenfunctions must be known.  For
this purpose we assume that the single-particle classical dynamics in
the nanoparticle is completely chaotic. In this case it is well-known
that the single-particle quantum mechanics is accurately described by
random matrix theory (RMT) {\em on an energy scale smaller than the
Thouless energy $\Eth$} (which, again, is the inverse time of flight
across the system) \cite{GutzBook,QChaosHouches}.  We consider only
the case of no symmetry here, so that the energy levels $\{ \epsilon_i
\}$ obey Gaussian Unitary Ensemble (GUE) statistics.  Because of the
spatial integrals in the expressions for $ M_{ij}$ and $N_{ij}$, the
correlations of the wavefunctions are also needed.  For the classical
Gaussian ensembles, there is no spatial correlation for a given
wavefunction, except for that arising from the
normalization. This is a direct consequence of the fact that no basis
should play a particular role for these matrix ensembles.  Here,
however, because of the particular role played by the kinetic energy,
the plane wave basis is special. More precisely, we should implement
that each eigenfunction is localized in this basis on the
energy scale $\Eth$.  We shall therefore build our
random matrix ensemble with the following requirements (a practical
implementation is given in section~\ref{sec:level2}):
\begin{itemize} 
\item[i)] On energy scales smaller than the Thouless
energy $\Eth$, one should recover standard GUE properties.  
\item[ii)]
On scales larger than $\Eth$, an eigenstate of energy $\epsilon$ should
appear localized on the energy surface $|\bk| = \sqrt{2m \epsilon}/\hbar$.
More precisely, we require that in a plane wave basis an eigenstate has
an envelope of width 
\begin{equation} \label{eq:delta_k}
           \delta k = 1/L
\end{equation} where $L$ is the typical size of the system.  
\item[iii)]
Finally, we have to implement that the fluctuations of
energy levels, and also importantly of local density, saturate at the
Thouless energy.  That is, if one considers a quantity $A$ expressed as
a sum over the energy levels in an energy window $\delta E$ about the
Fermi energy, as the window widens the fluctuations of $A$ increase until
$\delta E \!\approx\! \Eth$ beyond which point the fluctuations of $A$
do not further increase.  
\end{itemize}

Except for the fact that the Fermi surface is given an explicit width
$\approx \Eth$ due to the finite size $L$ of the system, items i)
and ii) are very similar to Berry's modeling of chaotic wave functions
\cite{BerryJ077} which assumes that a random superposition of plane waves
describes the correlation on scales smaller than the size of the system $L$.
Item iii) is motivated by the known saturation
of, for instance, fluctuations in the density of states of chaotic systems
at an energy scale of order $\hbar$ divided by the period of the shortest
periodic orbit \cite{GutzBook}.  Note that these requirements will yield
a random matrix approach noticeably different from the ones discussed in
Refs.~\cite{Prus96,BrouwOregHalp99,JacquStPRL00}.  

In this model, the wave function statistics that we find are as follows:
for a two dimensional dot of area $A$ (the case that we consider throughout 
this paper)
\begin{equation}\label{eq:psicor}
      A \langle \psi_i(\br) \psi_{j}^*(\br') \rangle =
          \delta_{ij}\, J_0 (\kf | \br - \br' |)
\end{equation}
for $|\br-\br'| \ll L$.  (In a maximum entropy approach semiclassically
restricted by the classical dynamics, this corresponds to keeping only the
direct trajectory between $\br$ and $\br'$; the approach can be generalized
to keep more classical paths \cite{Srednicki96,E2NBar01}.)  
In addition, it can be shown that within our random matrix
modeling, one has, up to $1/(\kf L)$ corrections
\begin{eqnarray}
  \lefteqn{\langle \psi_i(\br_1) \psi_{i}^*(\br_2) \psi_i(\br_3)
          \psi_{i}^*(\br_4)\rangle = } 
             \\ && \qquad
          \langle \psi_i(\br_1) \psi_{i}^*(\br_2) \rangle
	\langle \psi_i(\br_3) \psi_{i}^*(\br_4) \rangle \nonumber 
             \\ && \qquad
     + \langle \psi_i(\br_1) \psi_{i}^*(\br_4) \rangle
	\langle \psi_i(\br_3) \psi_{i}^*(\br_2) \rangle \label{eq:psicor2}\; .
           \nonumber
\end{eqnarray}
Alternatively,
these statistics can be obtained directly by assuming that the
distribution of wavefunctions is Gaussian \cite{Srednicki96}, though in this
case special care must be taken with regard to the normalization constraint. 

There are alternate routes to our very natural starting
point Eq.~(\ref{eq:total_E}).  One that has been
developed recently is the ``universal Hamiltonian'' approach
\cite{Agametal97,AleinGlaz98,MesoStoner00,AleinBrouwGlaz01} which uses RPA to
treat the interactions, RMT properties of the single-particle wavefunctions,
and the small parameter $1/g$ to arrive at an effective Hamiltonian. Treating
this effective Hamiltonian in the Hartree-Fock approximation leads to the
same expression for the ground state energy that we give above.

We end this section by giving explicit expressions for the
inverse compressibility, which is proportional to the peak spacing
[Eq.~(\ref{eq:sdef})], in two simple cases in which the standard up-down
filling of the states is assumed.  First consider the case of $N$ even with
$n \smeq N/2$. For the standard filling, it is a singlet state and the two
neighboring states for $N \!-\! 1$ and $N \!+\! 1$ electrons are doublets.
The peak spacing is then
   \begin{eqnarray}\label{eq:updown_evenspacing}
      \lefteqn{ \Delta^2E_{N\, {\rm even}} =
         \Delta^2 \ETF + \delta \epsilon_{n+1} - \delta \epsilon_{n} }
            \\ && \qquad
          + \sum_{i=1}^n \left[ 2 \bigl( M_{n+1,i} - M_{n,i} \bigr)
                              - \bigl( N_{n+1,i} - N_{n,i} \bigr) \right]
       \;. \nonumber
    \end{eqnarray}
On the other hand if $N$ is odd, still assuming the simplest up-down filling,
the peak spacing is
   \begin{equation}\label{eq:updown_oddspacing}
      \Delta^2E_{N+1\, {\rm odd}} =
            \Delta^2 \ETF + M_{n+1,n+1} \;.
    \end{equation}
The second difference of $\ETF$ is almost equal to the charging energy
$e^2/C$ but there is a small correction from the residual interactions;
this is discussed in detail below.

\section{Large Dot Limit}
   \label{sec:largedot}

The main simplifying feature in the large dot limit, $\kf L \gg 1$, is
that the fluctuations of the interactions can be neglected, much as for
diffusive nanoparticles~\cite{Agametal97,AleinGlaz98,Blant96,BlantMir97} in
the large $g$ limit.  The distribution of the spacings is determined, then,
by the fluctuations of the single-particle levels and their interplay with
the mean residual interactions.  Of course, as has been studied previously
\cite{Prus96,BrouwOregHalp99,BarUllGlaz00,MesoStoner00}, the ground state of the
nanoparticle does not necessarily follow the simple filling used in Eqs.
(\ref{eq:updown_evenspacing})-(\ref{eq:updown_oddspacing}), and this effect
must be included in finding the distribution.

Let us consider the mean and variance of the $M_{ij}$ and $N_{ij}$ for
levels near the Fermi level. We have, for instance, using
Eq.~(\ref{eq:psicor2}) 
\begin{eqnarray}\label{eq:avgMii}
      \lefteqn{ \langle M_{ij} \rangle =
         \int d\br d\br' \langle \left| \psi_i(\br) \right|^2 \rangle
            \Vtf(\br-\br') \langle \left| \psi_j(\br') \right|^2 \rangle  }
      \\ && \qquad \qquad
         + \delta_{ij} \int d\br d\br'
            \left| \langle \psi_i(\br) \psi_i^*(\br') \rangle \right|^2
               \Vtf(\br-\br') \nonumber
\end{eqnarray}
We will make the simplifying assumption that the nanoparticle is
a billiard so that the average density is constant: $\langle \left|
\psi(\br) \right|^2 \rangle \smeq 1/A$ where $A$ is the area of the
particle and the correlation function is given by Eq.~(\ref{eq:psicor}).
In this case the mean values are
  \begin{eqnarray} \label{eq:MN_av}
      \langle M_{ij} \rangle & = & \frac{\Delta}{2} +
         \delta_{ij} \frac{\langle \hV \rangle_{\rm fs}}{A} 
   \nonumber \\
      \langle N_{i \neq j} \rangle & = &
         \frac{\langle \hV \rangle_{\rm fs}}{A}
  \end{eqnarray}
where 
  \begin{equation}
	\langle \hV \rangle_{\rm fs} = 
	\int_0^{2\pi} \frac{ d\theta}{2\pi} \,
           \hV \bigl(\kf \sqrt{2 (1+\cos\theta)} \bigr)
  \end{equation}
is the average on the Fermi surface of $\hV ({\bf k} - {\bf k'})$.  In
the zero-range interaction limit, these expressions simplify to
  \begin{equation} \label{eq:MNlocal_av}
     \langle M_{ij} \rangle = (1+\delta_{ij}) \Delta/2 \; , \quad
        \langle N_{i \neq j} \rangle = \Delta/2 \; .
  \end{equation}
{\em The magnitude of the residual interaction effects is thus of order the
mean level spacing $\Delta$.} (Note that the $\Delta$ that we use here and
throughout this paper is the spacing of the orbital levels alone and so does
{\em not} take into account the spin degeneracy factor.) 
Since energies of this order
are critical in determining the spacing of CB peaks---note, for instance,
the case of standard up-down filling Eq.~(\ref{eq:updown_evenspacing})---the
residual interaction terms must be included in any theory. In particular,
the reason for the failure of the simple constant interaction model,
Eqs. (\ref{eq:CIenergy}) and (\ref{eq:CIprediction}), is now clear: it
does not consistently keep all terms of order $\Delta$.

The variance or covariance of any of the $M_{ij}$ and $N_{ij}$ will involve
the correlation function of a wavefunction. For example, consider the
variance of $M_{i \neq j}$:
\end{multicols}
\vspace*{-0.25truein} \noindent \hrulefill \hspace*{3.5truein}
\begin{eqnarray} \label{eq:M_var}
    {\rm var} ( M_{i \neq j} ) & = & \int \! d\br_1 d\br_2 d\br_3 d\br_4 
        \Bigl\langle [|\psi_i(\br_1)|^2 - A^{-1} ] \Vtf(\br_1 -\br_2)
                       [|\psi_j(\br_2)|^2 - A^{-1} ] 
        \\ && \qquad \qquad \qquad \times 
                     [|\psi_i(\br_3)|^2 - A^{-1} ] \Vtf(\br_3 -\br_4)
                       [|\psi_j(\br_4)|^2 - A^{-1} ] \Bigr\rangle
        \nonumber \\
         & = & \int \! d\br_1 d\br_2 d\br_3 d\br_4
            \Vtf(\br_1 -\br_2) \Vtf(\br_3 -\br_4)
          \left| \langle \psi_i(\br_1) \psi_i^*(\br_3) \rangle \right|^2
            \left| \langle \psi_j(\br_2) \psi_j^*(\br_4) \rangle \right|^2
       \; .
\end{eqnarray}
\vspace*{-0.15truein} \noindent \hspace*{3.5truein} \hrulefill 
\begin{multicols}{2}\noindent
The first equality is due to the normalization of the wavefunction,
and the second derives from Eq.~(\ref{eq:psicor2}).
Because of the decay of the autocorrelation of $\psi$, Eq.
(\ref{eq:psicor}), the spatial integrals in (\ref{eq:M_var}) are restricted
to a region where $\br_1$ is close to $\br_3$ and likewise $\br_2$
to $\br_4$. These restrictions produce factors of the small parameter
$1/\kf L$ compared to the mean values. We will evaluate these factors
below. However, for the large dot limit here, we see that to
zeroth order in $1/\kf L$ all the fluctuations in the residual interactions
may be neglected~\cite{BrouwOregHalp99,BarUllGlaz00,MesoStoner00}, as in
the diffusive case~\cite{Agametal97,AleinGlaz98,Blant96,BlantMir97}.
Qualitatively, this is natural because of the averaging
implicit in the integral in $M_{ij}$ and $N_{ij}$: when the size of the
system is much larger than the scale of oscillation of $\psi
(\br )$, particular features of $\psi_i$ and $\psi_j$ become less important
compared to the mean behavior.

For the CB peak spacing distribution, we thus consider the expression for
the ground state energy (\ref{eq:total_E}) with constant $\langle M_{ij}
\rangle$ and $\langle N_{ij} \rangle$.  The problem is still complicated
because of the interplay of the fluctuations in the single particle levels
with these constants, which may, for instance, lead to $S \smeq 1$ or higher
ground state spins \cite{BrouwOregHalp99,BarUllGlaz00,MesoStoner00}. We
thus evaluate the distribution of Coulomb blockade peak spacings numerically:
Hamiltonians are taken at random in the GUE ensemble, and the lowest energy
states for $N \!-\! 1$, $N$, and $N \!+\! 1$ particles are obtained by
determining in each case the occupation numbers $\{n_{i \sigma}\}$ which
minimize the expression Eq.~(\ref{eq:total_E}).

\begin{figure*}[t]
\begin{center}
\leavevmode
\epsfxsize = 7.3cm
\epsfbox{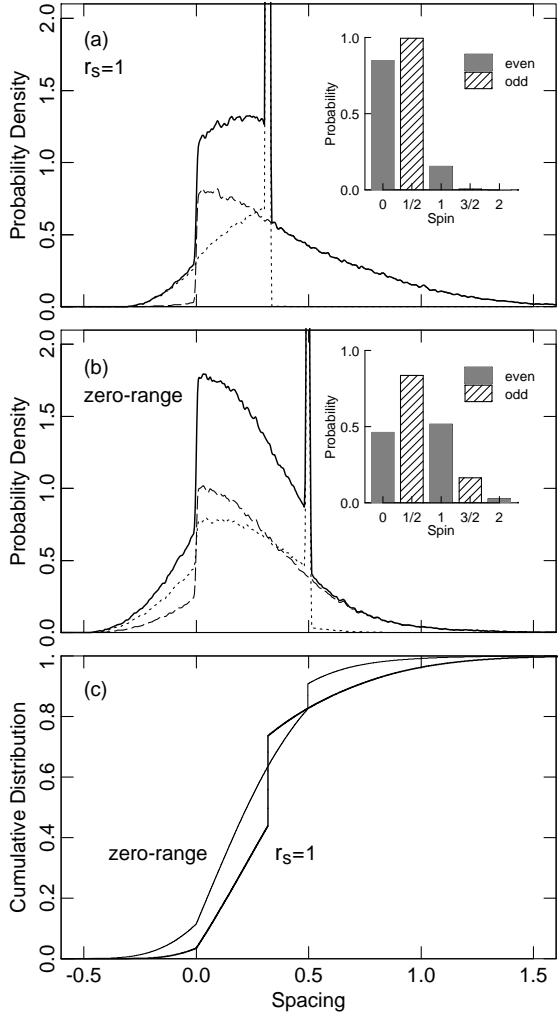}
\end{center}
\caption{
The probability density of CB peak spacings in the large dot limit ($\kf
L \rightarrow \infty$). The total distribution (solid) as well as that
for $N$ even (dashed) and odd (dotted) is given for two strengths of
the interactions: (a) $r_s \smeq 1$ and (b) the zero-range interaction limit.
The presence of a $\delta$-function in the distribution is particularly clear
in the cumulative distribution functions (the integral of the probability
density) shown in (c). Insets show the probability of occurrence of ground
state spins in the two cases.  The spacing is in units of the mean level
separation $\Delta$, and the origin corresponds to the classical spacing
$e^2/C$.
}
\label{fig1}
\end{figure*}

The resulting distribution of CB peak spacings is shown in Fig. 1.  We show
it for two different strengths of interactions, the case of $r_s \smeq 1$
and the zero-range interaction limit.  The first thing to note is that for both
of these values the distribution differs substantially from that in the CI
model, Eq.~(\ref{eq:CIprediction}), with much less odd/even alternation
present.  Still, there is a $\delta$-function in the distribution, as is
clear in the cumulative distribution functions in panel (c), because the
constant residual interaction term gives a rigid shift to the spacing for
odd $N$ and standard filling, Eq.~(\ref{eq:updown_oddspacing}).  The origin
of spacing in these plots is $\Delta^2 \ETF \!+\! \langle M_{i \neq
j} \rangle$. This is the mean spacing for adding a particle into the next
available state neglecting all spin/filling effects; it thus corresponds to 
the classical charging energy $e^2/C$.

The fluctuations of the peak spacings are {\it smaller} in this constant
exchange model than in the CI model. This is natural since the even and
odd parts of the distribution are brought closer together.  The values are
{\sl rms}$(\Delta^2 E_N )/\Delta \smeq 0.58$ in the CI model and $0.28$
($0.24$)
in the present large dot case for $r_s \smeq 1$ (zero-range interaction).

\begin{figure*}[b!]
\begin{center}
\leavevmode
\epsfxsize = 12cm
\epsfbox{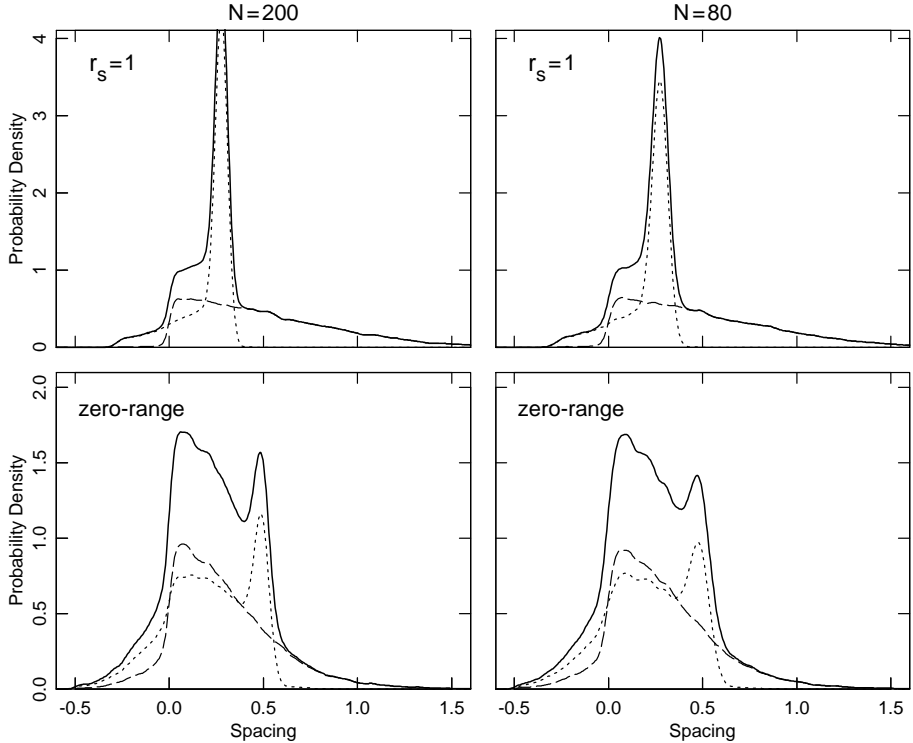}
\end{center}
\protect\caption{
The probability density of CB peak spacings including the fluctuations
in the residual interactions.  The total distribution (solid) as well
as that for $N$ even (dashed) and odd (dotted) is given for two sizes
of quantum dot ($N \!\approx\! 200$, left column, or $80$, right column) and
two strengths of interaction ($r_s \smeq 1$, top row, or the case of a
zero-range interaction, bottom row).  All of the features of the distribution
are smoothed compared to Fig. 1. Note that the odd/even
effect is quite clear in all four cases, and that the remnants of the
$\delta$-function are discernible.  The spacing is in units of the mean
level separation $\Delta$, and the origin corresponds to the classical
spacing $e^2/C$.
}
\label{fig2}
\end{figure*}

\end{multicols}
\vspace*{-0.25truein} 
\section{Fluctuations of Residual Interactions}
   \label{sec:residualint}

In order to evaluate the statistical effect of fluctuating residual
interactions, we need to find the distribution of the $M_{ij}$ and $N_{ij}$.
We will show that in the limit of large but not infinite $\kf L$, the
$M$'s and $N$'s are Gaussian distributed and uncorrelated.

The first step is to evaluate the variance of the $M$'s and $N$'s. 
Using    the    random    matrix    ensemble    introduced    in
Section~\ref{sec:approach}, we find to a good approximation, 
\begin{eqnarray} \label{eq:Mij_var}
     {\rm var} ( M_{i \neq j} ) & \simeq &
        \frac{8}{\pi A^2} \int_{\pi/L}^{2k-\pi/L} \frac{dq }{q}
           \frac{\Vtf(q)^2}{(2kL)^2 - (qL)^2} 
       \simeq 
        \frac{1}{\pi A^2} \frac{\ln(kL)}{(kL)^2}[2 \hV^2(0) + \hV^2(2k)]
                \\
     {\rm var} ( N_{i \neq j} ) & \simeq &
        \frac{2}{\pi A^2} \int_{\pi/L}^{2k-\pi/L} \frac{dq}{q}
           \frac{[\hV(q) + \hV(2k)]^2 }{(2kL)^2 - (qL)^2} 
                 \label{eq:N_var} \\
	\langle M_{i \neq j} N_{i \neq j} \rangle - 
	\langle M_{i \neq j}\rangle \langle N_{i \neq j} \rangle 
         & \simeq & 
        \frac{4}{\pi A^2} \int_{\pi/L}^{2k-\pi/L} \frac{dq}{q}	
           \frac{\hV(q)[\hV(q) + \hV(2k)] }{(2kL)^2 - (qL)^2} \;;
\end{eqnarray}
details for the case of ${\rm var} ( M_{i \neq j})$ are given in Appendix
\ref{sec:app_varMij}.  The diagonal part of the direct residual interaction
has an extra contribution because of the additional way in which the
wavefunctions can be paired:
   \begin{eqnarray} \label{eq:Mii_var}
      {\rm var}( M_{ii} ) \simeq
          2 {\rm var}( M_{i \neq j}^2 )
         + \frac{8}{\pi A^2} \int_{\pi}^{2k-\pi/L} \frac{dq}{q}
           \frac{\hV(q) [\hV(0)+\hV(\sqrt{(2k)^2 - q^2})]}{(2kL)^2 -(qL)^2}
        \; . 
   \end{eqnarray}
\vspace*{-0.10truein} \noindent \hspace*{3.5truein} \hrulefill
\begin{multicols}{2}\noindent

In the zero-range interaction limit, the expressions for the variance of the
$M$'s and $N$'s simplify considerably, as for the mean values given in
Eq.~(\ref{eq:MNlocal_av}).  In this case, we find
   \begin{equation} \label{eq:sr_var}
      {\rm var} ( M_{ij} ) = {\rm var} ( N_{i \neq j} )=
     \frac{3 \Delta^2}{4\pi} \frac{\ln(kL)}{(kL)^2} (1+3\delta_{ij})
        \; .
   \end{equation}
Note that the decay of the wavefunction correlations appearing in the
variance produces a factor of $1/\kf L$ in the rms compared to the mean.
The $\ln(kL)$ factor is special to two dimensions; it comes from the
$1/kL$ decay of the wavefunction correlator in this case.  Thus in the
large dot limit, $\kf L \!\gg\! 1$, the distributions of the $M$'s and $N$'s
are narrowly peaked about their mean values. This is the justification
for ignoring these fluctuations to a first approximation in the last section.

Higher moments of the distribution involve coupling between more pairs
of wavefunctions.  The non-Gaussian part of the distribution is described
by the cumulants, in which all of the wavefunctions are coupled in ways
not present in the lower moments. In this case, the spatial integrals
will have more restrictions coming from the wavefunction correlation
(\ref{eq:psicor}); these restrictions will in turn produce additional factors
of the small parameter $1/\kf L$.  Thus in the large dot limit all higher
moments can be neglected. In the case of third and fourth moments this can
be easily verified by explicit calculation.  Likewise it is straightforward
to check that cross correlations among the $M$'s and $N$'s do not need
to be considered, except for the obvious correlation between $M_{ij}$ and
$N_{ij}$ which are equal in the zero-range interaction limit.  We shall return
to this point in Section~\ref{sec:level2}.  In this Section, however,
we shall either take $M_{ij} \smeq N_{ij}$ in the zero-range interaction limit, 
or
neglect this correlation for $r_s \smeq 1$, and therefore take the $M_{ij}$
and $N_{ij}$ to be uncorrelated random Gaussian variables with mean and
variance given above.


We  now use  the  statistics  of the  $M_{ij}$,  $N_{ij}$, and  single
particle levels (assumed  to follow the random matrix  GUE ensemble as
in Section \ref{sec:largedot}) to deduce the statistics of the CB peak
spacings.  It  is clear  from the expressions  for the spacing  in the
simple up-down  filling case, Eq.~(\ref{eq:updown_evenspacing}), that
sums  such  as $\sum_j  M_{ij}$  will  be  important. Because  of  the
saturation of the  fluctuations, item iii) in the  random matrix model
of  Section \ref{sec:approach}, the  variance of  such a  sum involves
only levels  within $\Eth$  of each other.  In calculating  ${\rm var}
\bigl( \sum_{j} M_{ij} \bigr)$, then,  the sum should be taken over of
order $\kf L$ independent $M$'s:
\begin{equation} \label{eq:sumM_var}
      {\rm var} \Bigl( \sum_{j} M_{ij} \Bigr)
          \propto \frac{\ln(\kf L)}{\kf L} \Delta^2 \; .
\end{equation}
The variance of other sums over the residual interaction terms will
have the same dependence on $\kf L$.  {\em Up to a $\sqrt{\ln}$
factor, we thus expect the contribution of the fluctuations of the
residual interactions to the CB peak spacings to be of order $\Delta
/\sqrt{\kf L}$ in rms.}

In order to evaluate the magnitude more accurately, we turn again to
numerical evaluation.  We consider a collection of $\Eth/\Delta$ levels
with GUE energies and Gaussian residual interaction terms. We fill the
levels with $N \!-\! 1$, $N$, and $N \!+\!  1$ particles successively,
and find the occupation numbers $\{n_{i\sigma}\}$ minimizing the
energy Eq.~(\ref{eq:total_E}) in each case.  Fig. 2 shows the resulting
distribution of CB peak spacings.  The fluctuations clearly act to smooth
out all the sharp features of the $\kf L \rightarrow \infty$ case: the
$\delta$-function becomes a finite peak, and the discontinuity at the origin
is removed.  The difference between the $N$ even and odd cases is quite
clear---even for the relatively small $N \smeq 80$ (for which $1/\sqrt{\kf
L} \!\simeq\! 0.21$).  The fluctuations of the residual interaction terms
lead, of course, to an increase in the fluctuations of the spacing; now
{\sl rms}$(\Delta^2 E_N )/\Delta \smeq 0.31$ for $r_s \smeq 1$ and $N
\smeq 80$-$200$ compared with $0.28$ in the absence of such fluctuations.
While the inclusion of the fluctuations has clearly made the distributions
more like the experiments, substantial differences remain.


\section{Scrambling: Added Electron Changes Confinement}
   \label{sec:scrambling}

As electrons are added to the nanoparticle, the mean field potential which
confines the electrons changes.  This causes a change in the single particle
levels, whose energies appear in the expression for the ground state energy
Eq.~(\ref{eq:total_E}), which then in turn change the CB peak spacing.
Because we consider a chaotic quantum dot, the change in the energy
levels will be unpredictable; hence, we call this effect ``scrambling''.
Note that it is an intrinsic effect connected to the charge of the electron
not the geometry or environment of the dot.  This effect has been studied
previously for diffusive quantum dots in Ref. \onlinecite{BlantMirMuz97};
here we give a derivation directly in the context of ballistic chaotic dots.

We know from the Strutinsky approach (Section \ref{sec:approach}) that
the change in the single particle levels should be found from the change
in the smooth Thomas-Fermi potential, Eq.~(\ref{eq:deltaE1}).  So we first
evaluate this change in potential, and then use perturbation theory to find
the change in the energy levels.  Perturbation theory is justified since
the resulting shift is smaller than the mean level separation $\Delta$ by
a factor depending on $\kf L$.  As a check, one can evaluate the magnitude
of the fluctuation in the event of a complete scrambling of all the single
particle levels. The properties of sums of random matrix eigenvalues in
Ref. \onlinecite{Leboeuf00} 
imply a very large fluctuation of the peak spacings---of
the order $2$-$4 \Delta$ which is larger than observed experimentally.

\subsection{Evaluation of ${\mathbf \delta \Veff}$}

The generalized Thomas-Fermi problem is specified in terms of the density
functional
 \begin{equation}
   \FTF[n] = \TTF[n] + \Eext[n] + \Ecoul[n] + \Exc[n]
 \end{equation}
comprising four contributions: the kinetic energy
 \begin{equation}
   \TTF[n] = \int \!\! d\br \Bigl[ \int^{n(\br)} \!\!dn'
                 \epsilon(n') \Bigr]
 \end{equation}
where $\epsilon(n')$ is the maximum kinetic energy of non-interacting
particles with density $n'$, the confinement potential defining the
geometry of the dot
 \begin{equation}
   \Eext[n] = \int \!\! d\br \, \Vext(\br) n(\br) \;,
 \end{equation}
the Coulomb energy
 \begin{equation}
   \Ecoul[n] =
     \frac{e^2}{2}\int \!\! d\br d\br' \frac{n(\br)n(\br')}{|\br-\br'|} \;,
 \end{equation}
and the exchange-correlation potential $\Exc$.  The Thomas-Fermi density
is found by solving
 \begin{equation} \label{eq:TF}
    \frac{\delta \FTF}{\delta n}[\nTF] = \mu_{TF} \; ,
 \end{equation}
where $n(\br)$ is the electronic density and $\mu_{TF}$ is found such that 
$\int n(\br) d\br = N$.  The effective potential is, then, defined by
 \begin{equation}
   \Veff(\br) \equiv
      \frac{\delta (\Eext + \Ecoul + \Exc )}{\delta n}[\nTF] \;.
 \end{equation}

Let $\delta \nTF = \nTF(N+1) - \nTF(N)$
be the change in the Thomas-Fermi density when one electron is added to
the system.  Bearing in mind that Eq.~(\ref{eq:TF}) is a classical-like
equation for which $\delta \nTF$ is small, one can write
   \begin{eqnarray} \label{eq:TFlin}
      \lefteqn{
	\int \! d\br' \frac{\delta^2 \TTF}{\delta n^2}[\nTF](\br,\br')
	    \delta \nTF(\br') 
      } \\ && \qquad \qquad  \qquad 
         + \int d\br' \frac{\delta \Veff }{\delta n}(\br,\br')
	    \delta \nTF(\br') = \delta \mu_{TF} \;
      \nonumber
   \end{eqnarray}
with $\delta \mu_{TF}$ such that $\int \delta \nTF(\br)=1$.  
In two dimensions, the Coulomb energy of the added charge will dominate
any variations in the kinetic energy.  One can
therefore write the variation of the density as $ \delta \nTF = \delta
\nTF^0 + \delta \nTF^1 + \ldots$ with $\delta \nTF^0$ the solution of the
electrostatic problem
   \begin{eqnarray}
	\delta \mu^0 & = & \int d\br'
	   \frac{\delta^2 \Ecoul}{\delta n^2}(\br,\br') \delta
	      \nTF^0(\br')
      \nonumber \\
	& = & e^2 \int d\br' \frac{\delta \nTF^0(\br')}{|\br-\br'|} \; .
   \end{eqnarray}

At this level of approximation, the variation of the potential {\em inside}
the dot is just the constant $\delta \mu^0$, which will just shift the one
particle energies $\e_i$ but not give rise to any fluctuation.  One should
keep in mind though that Eq.~(\ref{eq:TF}) applies only at places where
$\nTF(\br) \neq 0$.  Therefore the {\em boundaries} of the effective
potential can be affected, in a way that depends, of course, largely on the
external confining potential, and in particular on the second derivative
of this latter at the the boundary.  If the curvature is weak, fairly large
displacements of the boundary can occur, which could play a significant role.

If, on the other hand, one assumes a billiard-like confining potential,
the only source of modification of the effective potential comes from
$\delta \nTF^1$, which is obtained through the equation
   \begin{equation}
	\int d\br' \frac{\delta^2 \TTF}{\delta n^2}(\br,\br') \delta
	\nTF^0(\br') + \int d\br' \frac{\delta \Veff }{\delta n}(\br,\br')
	\delta \nTF^1 = \delta \mu^1 \; .
   \end{equation}
Noting that $\int (\delta \Veff/ \delta n) \delta \nTF^1$ is, up to
an inessential additive constant, the variation $\delta \Veff$ of the
effective potential we are interested in, and that
   \begin{equation}
      \frac{\delta^2 \TTF}{\delta n^2}(\br,\br')
	 = \nu^{-1} \delta_d (\br-\br') \;,
   \end{equation}
we get
   \begin{equation} \label{eq:deltaV}
      \delta \Veff(\br)
	= - \Delta \, \bigl( \delta \tilde n (\br)
			 - \overline{\delta \tilde n } \bigr)
   \end{equation}
where $\delta \tilde n \equiv A \delta \nTF^0(\br) $ is a smooth function
of order $1$ ($A$ is the area of the dot). Note that we choose $\delta
\Veff$ with mean value zero. This ensures that there is no change in the
mean energy levels; any such change in the mean should be incorporated in
the charging energy.

\subsection{Consequences for the peak spacing distribution}

The variation of the effective potential induces a change of the one particle
energies. We let $\ve_N$ denote the single particle energy of the
$N^{\mbox{th}}$ electron, whatever orbital it is in (for $N$ even and simple
up/down filling, for instance, $\ve_N \smeq \epsilon_{N/2}$).
In perturbation theory,
$\delta \ve_j = \delta\ve^{(1)}_j + \delta\ve^{(2)}_j + \ldots$ with
  \begin{eqnarray} \label{eq:1storder} 
     \delta\ve^{(1)}_j & = & \langle \psi_j | \delta \Veff | \psi_j \rangle \\
        \delta\ve^{(2)}_j & = & \sum_{k\neq j}
           \frac{|\langle \psi_j | \delta \Veff | \psi_k |
     \rangle|^2}{\ve_k - \ve_j}
	\label{eq:2ndorder}\; .
  \end{eqnarray}
When going back to the inverse compressibility, one should bear in mind,
however, that the variation of $\Veff$ when going from $N \!- \!1$ to $N$ electron
is the same as when going from $N$ to $N \!+ \!1$.  As a consequence, it can be
checked easily that the linear variation $\delta\ve^{(1)}_j$ cancels for
all levels except the two top ones.  At this level of approximation one
therefore gets
  \begin{equation}
     \Delta^2E_N = \Delta^2E_N^{\rm CI} + \Delta^2E_N^{\rm scramb} \;
  \end{equation}
where $\Delta^2E_N^{\rm CI}$ is the constant interaction model result given 
in Eq.~(\ref{eq:CIprediction}), and
  \begin{equation} \label{eq:Del2Escramb}
    \Delta^2E_N^{\rm scramb} =
	\delta \ve_{N+1} + \delta \ve_N + \sum_{j=1}^{N-1} \delta^2 \ve_j
  \end{equation}
is the correction due to scrambling. The levels used for $\Delta^2E_N^{\rm CI}$ 
are those for the $N$-electron $\Veff$.  Throughout this scrambling argument
we assume that the filling of the levels does not change.

\begin{figure*}[b]
\begin{center}
\leavevmode
\epsfxsize = 12cm
\epsfbox{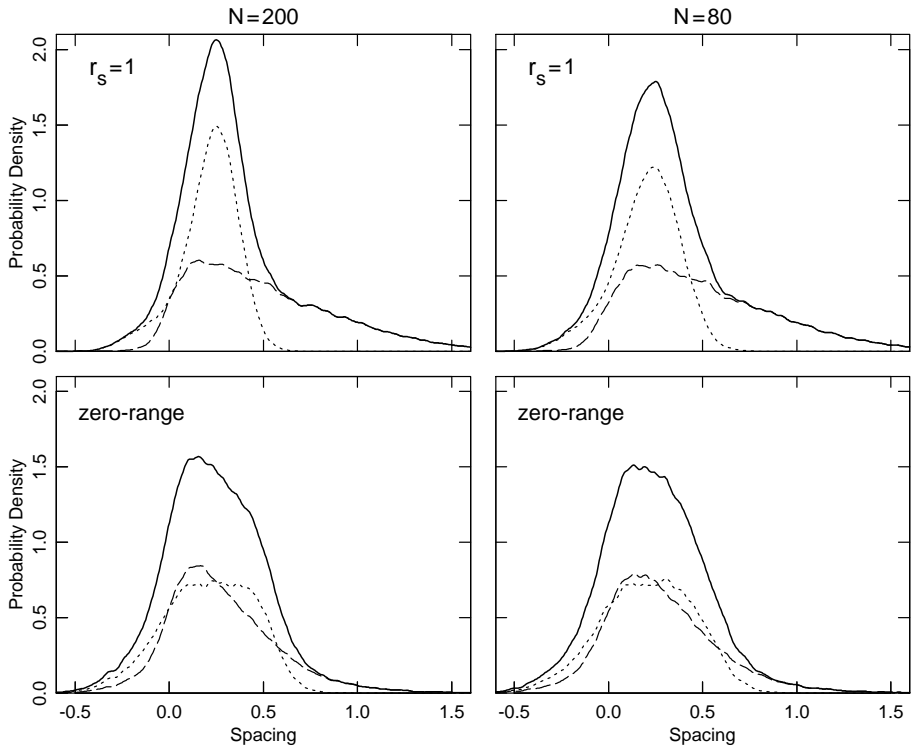}
\end{center}
\protect\caption{
The probability density of CB peak spacings including the scrambling
effect on the top two levels as well as the fluctuations in the residual
interactions.  The total distribution (solid) as well as that for $N$ even
(dashed) and odd (dotted) is given for two sizes of quantum dot ($N \smeq
200$, left column, or $80$, right column) and two strengths of interaction
($r_s \smeq 1$, top row, or the case of a zero-range interaction, bottom row).
Note that the clear odd/even effect for $r_s \smeq 1$ and the relatively
small $N \smeq 80$ remains visible even for the stronger zero-range interaction.
The spacing is in units of the mean level separation $\Delta$, and the
origin corresponds to the classical spacing $e^2/C$.
}
\label{fig3}
\end{figure*}

\subsubsection*{First order terms}

The mean of the first order terms is zero by construction.  The variance
of $\delta\ve^{(1)}_i$ is
   \begin{equation} \label{eq:var1}
      {\rm var} \bigl(\delta\ve^{(1)}_i\bigr) = \Delta^2 \int \!\! d\br d\br'
            |\langle \psi_i(\br) \psi_i^*(\br') \rangle |^2
      \delta \tilde n (\br) \delta \tilde n (\br') \;.
   \end{equation}
Using the wavefunction correlations (\ref{eq:psicor})
and approximating $J_0^2(\kf|\br-\br'|) \simeq 1 / (\pi \kf |\br-\br'|)$
plus oscillating terms ($\kf |\br-\br'| > 1$), we find
 \begin{equation} \label{eq:var1a}
   {\rm var}\bigl(\delta \ve_i\bigr) \simeq \alpha \cdot \frac{\Delta^2}{\pi\kf L}
      \;.
 \end{equation}
The prefactor depends on the geometry of the dot under consideration:
 \begin{equation}
   \alpha = \int d^2\bu d^2\bu' \frac{1}{|\bu-\bu'|}
    \delta \tilde n(\bu) \delta \tilde n(\bu')
 \end{equation}
is a dimensionless quantity of order 1 (note $\bu = \br/L$).  For a circular
dot, for instance, with diffuse boundary scattering to make it chaotic, 
$\delta \tilde n = \{ [1-(r/R)^2]^{-1/2} -2 \} /2$ where $R$
is the radius; in this case, $\alpha \simeq 1.2 $.

Bearing in mind that $\kf L \simeq \sqrt{2\pi N}$ (taking into account the
spin degeneracy), we see that the typical amplitude of the variation of the
$\delta \ve_i$ scales as $N^{-1/4}$ times the mean level spacing.  This may,
in fact, be numerically significant for experimental dots---for a 
circular dot with 100 electrons, for instance, the magnitude would be
$\simeq 0.12 \Delta$ .

\subsubsection*{Second order terms}

The variance of the sum $\sum_1^{N-1} \delta^2 \ve_i$ can be evaluated
along the same lines.  Since it turns out to be parametrically smaller, 
we will be less careful with the constant. 

First we note that the fluctuation of an individual matrix element is
 \begin{equation}
   {\rm var} \bigl(\langle i | \delta \Veff | j | \rangle \bigr)  =
      \frac{\Delta^2}{A^2} 
           \int \!\! d\br d\br' J_0(k_i\delta \br) J_0(k_j \delta \br)
   \delta \tilde n(\br) \delta \tilde n(\br') 
 \end{equation}
where $\delta \br = \br-\br'$.
Following the same reasoning as above, this implies that
 \begin{equation}
   {\rm var} \bigl(\langle i | \delta \Veff | j | \rangle \bigr) \simeq
 \left\{ 
    \begin{array}{ccc}
         0 & {\rm if} & \delta k L > 1\\
         \alpha \Delta^2 /\pi \kf L & {\rm if} & \delta k L < 1
    \end{array} \right.
 \end{equation}
where $\delta k = k_i - k_j$.

This result is now inserted in the expression (\ref{eq:2ndorder}) for the
energy shift $\delta \ve_j^{(2)}$ and then for the inverse compressibility
Eq.~(\ref{eq:Del2Escramb}). Assuming the matrix elements $\langle i | \delta
\Veff | j | \rangle$ are Gaussian distributed independent variables, we find


 \begin{eqnarray}
  \sigma_2^2 & \equiv & {\rm var} \left(\sum_{i=1}^{N-1} \delta^2 \ve_i \right)
        \\
     & \simeq & \sum_{i=N-N_c}^{N-1} \sum_{j=N}^{N+N_c}
 \frac{1}{(\bar \ve_i- \bar \ve_j)^2} {\rm var}\bigl(
              |\langle \psi_i | \delta \Veff | \psi_j | \rangle|^2
      \bigr) \;.
   \nonumber
 \end{eqnarray}
Because of the saturation of the fluctuations of the levels, point (iii)
in the random matrix criteria of Section \ref{sec:approach}, the sum over
levels is cut off at $N_c \smeq \kf L$.  The final result is
 \begin{equation}
   \sigma_2^2 \propto \ln \kf L \left(\frac{\Delta}{\kf L}\right)^2
 \end{equation}
which is parametrically smaller than the first order result (\ref{eq:var1a}).
We therefore see that the the effect of scrambling is dominated by the
variation of the top levels, and that this adds contributions to the peak
spacing fluctuations which while parametrically small on the scale of
$\Delta$ could be numerically important for dots which are not too large.

\subsection{Numerics}

It is a simple matter to incorporate the scrambling of the top two levels
into our numerical investigation of the ground state energy specified
by Eq.~(\ref{eq:total_E}).  In random matrix theory, the change in
an energy level upon varying a parameter is uncorrelated with its
value \cite{Wilkinson89,WilkinsonAustin92}, so we simply choose random
$\delta\ve_N$ and $\delta\ve_{N+1}$ from a Gaussian distribution with variance
Eq.~(\ref{eq:var1a}). The ground state configuration is then found as before.

Fig. 3 shows the resulting probability density of CB peak spacings.  In the
$r_s \smeq 1$ case there is still a substantial difference between the
distributions of even and odd spacings: for $N$ even, the large spacing
tail coming from the level-spacing distribution (Wigner surmise) survives
even for less than 100 electrons.  When the interactions are stronger,
as in the zero-range interaction case shown, the even and odd distributions
are more similar, but there is still a discernible difference.
The magnitude of the fluctuations increases further: we now find
{\sl rms}$(\Delta^2 E_N )/\Delta \smeq 0.33$-$0.32$ for $r_s \smeq 1$ 
and $N \smeq 80$-$200$.


\section{Beyond the Gaussian model}
   \label{sec:level2}

Our treatment of the fluctuations of the residual interactions in the
previous sections relies on a certain number of assumptions.  We have
for instance neglected correlations among the $M_{ij}$ and $N_{i'j'}$
(except for the obvious $M_{ij} \smeq N_{ij}$ for a zero-range
interaction) when we know, for instance, that the correlation coefficient
of $M_{ij}$ with $N_{ij}$ is of order 1 even in the infinite size
limit.  Similarly we have assumed the absence of correlation between
these residual interaction terms and the one particle energies
$\e_i$. Furthermore, we have used expressions such as
Eqs.~(\ref{eq:Mij_var})-(\ref{eq:Mii_var}) which are derived in the
large $\kf L$ limit.  To check that these assumptions do not
drastically modify the resulting distribution of CB peak spacings, we
shall in this section implement the random matrix ensemble described
in Section~\ref{sec:approach} [paragraph containing
Eq.~(\ref{eq:delta_k})] by directly performing numerical Monte Carlo
calculations.


\begin{figure*}[tb]
\begin{center}
\leavevmode
\epsfxsize = 12cm
\epsfbox{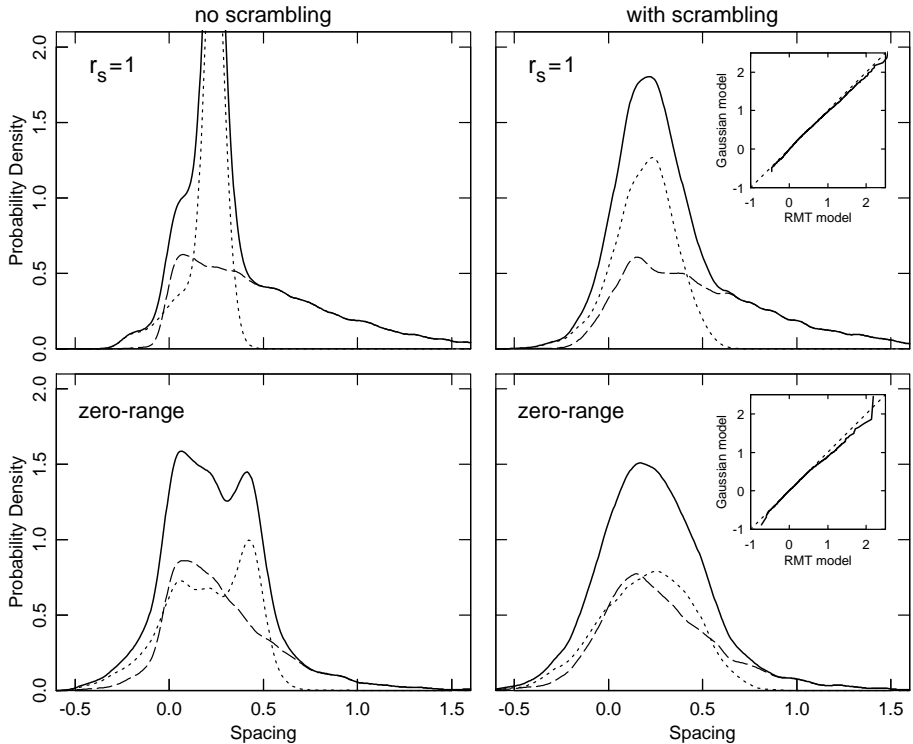}
\end{center}
\protect\caption{ The probability density of CB peak spacings within a
random matrix model which goes beyond the Gaussian approximation.  For
$N \!\approx\! 80$, the total distribution (solid) as well as that for
$N$  even (dashed)  and odd  (dotted) is  given for  two  strengths of
interaction  ($r_s \smeq  1$, top  row, or  the case  of  a zero-range
interaction, bottom row) in  the presence of scrambling (right column)
and without  scrambling (left  column).  Features of  the distribution
are  more pronounced  than  in the  Gaussian  model (Figs.  2 and  3),
suggesting that  smoothing effects  are slightly overestimated  in the
Gaussian approximation.  Insets: Quantile-quantile plots of the random
matrix model versus the Gaussian  model; note the differences at large
spacings.   Throughout, the  spacing is  in  units of  the mean  level
separation  $\Delta$,  and the  origin  corresponds  to the  classical
spacing $e^2/C$.  }
\label{fig4}
\end{figure*}
 
To implement the constraints of our random matrix model, we start by
considering the system to be a square of size $L$, for which we can use
a plane wave basis
 \begin{equation} \label{eq:planewave}
    |n_x,n_y\rangle = L^{-2} \exp[i 2\pi n_x x /L]\exp[i 2\pi n_y y /L]
 \end{equation}
ordered by increasing momentum.  In this basis, we write the one particle
Hamiltonian as $H \smeq D \!+\! V$ where $D$ is diagonal with $D_{ii}=i$.
$V$ is taken to be a banded random Hamiltonian with independent complex
matrix elements which are Gaussian distributed with variance $v^2_{ij} =
a(i) exp(-b(i) |i-j|)$.  For $a(i)$ in the correct range \cite{FyodMirl91}
such an ensemble yields (1)~eigenvalues with a mean spacing close to one,
(2)~local fluctuations which are typical of the GUE, and (3)~eigenfunctions
which are localized in energy.  The localization in energy can be
characterized by the inverse participation ratio in the plane wave
basis: $I_j = \sum_{n_x,n_y} |\langle n_x,n_y | \epsilon_j\rangle|^4$.
Our requirement Eq.~(\ref{eq:delta_k}) implies the constraint
   \begin{equation} \label{eq:IPR}
      I_j^{-1} = (3/4) (k_jL)\; .
   \end{equation}
In practice, we use a simple algorithm to tune the value of $a(i)$ and
$b(i)$ in such a way that this constraint holds.  Diagonalizing the random
matrix specified leads to eigenvalues and eigenvectors which are then
used to evaluate ground state energies for different number of particles
via Eq.~(\ref{eq:total_E}), and hence arrive at the CB peak spacings in
this model.

In  this  approach,  deviations  from  the  wavefunction  correlations
(\ref{eq:psicor})    required   by    normalization    are   naturally
present.    Furthermore,   correlation    between    eigenvalues   and
eigenfunctions  is present  as well.  Of  course, in  a real  physical
system, these  effects would be automatically included  because of the
scattering at  the boundary of  the system---the transit  time $\tau$,
for instance,  gives both the  scale over which  fluctuations saturate
($\hbar  /\tau$) and  the length  scale at  which the  Bessel function
correlations become suspect ($v_F  \tau$).  We believe that our random
matrix model includes the main  features of these effects in a generic
way.

Fig. 4 shows numerical results for this model. We consider only the smaller
value of $N$, $N \smeq 80$, because the effects are expected to be only
significant for smaller quantum dots. On the other hand, we present results
both with and without the scrambling effect of Section \ref{sec:scrambling},
and for different interaction strengths---$r_s \smeq 1$ corresponds to
a typical value for a GaAs quantum dot while the zero-range interaction case
corresponds to the maximum effect possible within a perturbative approach to
the interactions. When these results are compared to those of Figs. 2 and 3,
we see that there is a small but noticeable effect.  This is highlighted in
the quantile-quantile plots in the insets to Fig. 4 where an ordered set
of RMT spacings is plotted against a similar set for the Gaussian model.
The present RMT yields a more pronounced large spacing tail than the
Gaussian model.


\section{Conclusions}
   \label{sec:conclu}

In this paper we have studied the Coulomb blockade peak spacings for large
but not infinite quantum dots in the case that the interactions can be
treated perturbatively and the single-particle properties described by
random matrix theory.  Motivated by the Strutinsky approach to finite
Fermi systems \cite{Strutinsky68,Strutinsky72,UllTatTomBar01}, we use
an expression for the ground state energy [Eq.~(\ref{eq:total_E})] which
includes the charging energy, the single particle quantization effects, and
the residual (screened) interactions treated in first order (Hartree-Fock).

We presented results in four cases. First, in the limit $\kf L \!\rightarrow\!
\infty$, fluctuations of the residual interactions are negligible.
The CB peak spacing distribution has a discontinuity at $e^2/C$ and a
$\delta$-function because of the odd $N$ case (Fig. 1). The distribution
is very different from that predicted by the simple CI model which neglects
some terms of order $\Delta$ while keeping others.

Second, fluctuations in the residual interactions contribute a term
typically of order $\Delta/\sqrt{\kf L}$ to the CB peak spacings.
The sharp features of the infinite dot limit are rounded out (Fig. 2).

Third, the addition of an electron in the CB process changes the
confinement potential in the quantum dot and so changes the single particle
properties. This scrambling effect also contributes a term typically of
order $\Delta/\sqrt{\kf L}$ to the CB peak spacings, and the
distribution is rounded further. However, the difference between $N$
even and $N$ odd remains substantial at large spacings where a tail is
present for $N$ even (Fig. 3) which is a remnant of the Wigner surmise
distribution for level spacings.

Finally, we took a first step towards investigating corrections to
the Gaussian uncorrelated model for the fluctuations of the residual
interactions.  We found that the large spacing part of the distribution
is somewhat underestimated in the Gaussian model.

Before ending this paper, we compare our
theoretical distributions to those obtained experimentally
\cite{Sivan96,Simmel97,MarcusPatel98,Simmel99,LuschEnsslin01}.  The observed
distribution has little structure, being Gaussian near its peak but with
longer tails for both larger and smaller spacings.  The width of the
distribution is contentious: early work claimed that the width scaled with
the charging energy \cite{Sivan96,Simmel97} but this has not been verified
by more recent experiments \cite{MarcusPatel98,Simmel99,LuschEnsslin01}.
In the more recent experiments, the width is clearly related to the mean
level separation $\Delta$ and in the range $0.5$-$1.5 \Delta$, 
somewhat larger than given by, e.g., the CI model.

The theoretical results presented here certainly look more like the
experiments than the benchmark CI model: the distribution is more
rounded and so less structured, and the odd/even alternation is much less
pronounced. We suspect that further development of this approach will
produce {\it qualitative} agreement with the experiments.

In the end, however, the theory presented here does {\em not} provide
an adequate {\it quantitative} description of the experiments.  To
start with, the magnitude of the fluctuations is actually less than in
the CI model (0.58 versus 0.32 for $N \smeq 200$) and so differs from experiment. 
With
regard to the odd/even effect, the situation is somewhat less clear,
and so more interesting.  Looking at Fig.~3, for instance, one
observes that a substantial odd/even effect is still present for
$r_s=1$, while it has been largely washed out in the zero-range limit
of the screened interaction.  It should be stressed that this drastic
modification of the distributions can be traced back mainly to the
relatively modest change in the mean values $\langle M_{ii} \rangle -
\Delta/2$ and $\langle N_{ij} \rangle$,
Eqs.~(\ref{eq:MN_av})-(\ref{eq:MNlocal_av}), which change from about $0.3
\Delta$ for $r_s \smeq 1$ to $0.5 \Delta$ in the zero-range limit.  The
sensitivity to the numerical value of these quantities should not come
as a surprise since we know that the Stoner instability, for which a
spontaneous polarization of the spin of many electrons takes place,
occurs for $\langle N_{ij} \rangle = \Delta$.  These observations
suggest an experiment: placing a good conductor close to the 2DEG in
such a way that it can screen the interaction between electrons within
the quantum dot should significantly affect the peak spacing
distribution.  If the auxiliary screening is strong enough, a strong
odd/even effect should be visible. A double quantum well structure
seems like a possible setting for such an experiment.

We can divide the potential explanations for the quantitative discrepancy
between theory and experiment into basically three categories.  The first
one, which we cannot absolutely exclude, is that noise in the experiments is
corrupting the results \cite{MarcusPatel98,Heinzel_private}.  The second is
that effects not included here, but interpretable within the kind of Fermi
liquid theory on which we have relied throughout this paper, dominate.
Some candidates are (a) the effect of the changing gate potential on the
single particle levels \cite{Vallejos98,Vallejos99}---this would, however,
somewhat contradict the remarkable stability observed in the 
wavefunction \cite{MarcusStewart97}---(b) boundary effects in the
scrambling which need to be considered using a realistic description of
the dot, and (c) consequences of non-chaotic dynamics within the 
dots~\cite{Stopa98,Vallejos99}. 
Finally, a third completely different kind of
interpretation is that the perturbative approach to interactions taken here
is not accurate even though 
$r_s \!\approx\!  1$-$2$~\cite{Sivan96,Berkovits98,Prus96,Berkovits97}.
Further experimental and theoretical work is needed to discern between
these options.

\section*{Acknowledgements}
We thank G. Usaj for many valuable conversations, and appreciate helpful
discussions with I. L. Aleiner, L. I. Glazman, and C. M. Marcus.  The LPTMS
is a ``Unit\'e de recherche de l'Universit\'e Paris 11 associ\'ee au
C.N.R.S.''

\appendix

\section{Variance of the $M_{ij}$}
\label{sec:app_varMij}

In this appendix we sketch the argument which leads to
Eqs.~(\ref{eq:Mij_var})-(\ref{eq:Mii_var}).  As in
Section~\ref{sec:level2}, we assume that the configuration space is a
square of size $L$, for which the plane wave basis $|\bk\rangle =
|n_x,n_y\rangle $ introduced in Eq.~(\ref{eq:planewave}) can be used.

Expanding the eigenfunctions  in this basis,
\begin{equation}
 | \varphi_i \rangle = \sum_\bk v_{i\bk} |\bk\rangle \; ,
\end{equation}
we can write the $M_{ij}$'s as 
\begin{eqnarray}
	M_{ij} & = & \sum_{\bk_1,\bk_2,\bk_3,\bk_4}
	\!\!\! v_{i\bk_1} v^*_{i\bk_2} v_{j\bk_3} v^*_{j\bk_4}
	\delta_{\bk_1-\bk_2,-\bk_3+\bk_4} \hV(\bk_1-\bk_2) 
            \nonumber \\
	 & = & \sum_{\bq} \hV(\bq) W_{i \bq} W^*_{j \bq}
\end{eqnarray}
where we define
\begin{equation}
	W_{i \bq} \equiv \sum_{\bk_1-\bk_2=\bq} v_{i\bk_1} v^*_{i\bk_2} \; .
\end{equation}

Our random matrix model (Section \ref{sec:approach}) implies
\begin{eqnarray}
	\langle v_{i \bk_1} v^*_{j\bk_2} \rangle & = & (\kf L)^{-1} \delta_{ij}
	\delta_{\bk_1 \bk_2} 	\qquad \mbox{ if }  \delta k < \pi/L \\
	& = & 0 \; 	\qquad\qquad \qquad \qquad \mbox{ if }  
	\delta k > \pi/L
        \nonumber
\end{eqnarray}
with $\delta k = \left| |\bk_1| - k_i \right|$ and 
$k_i = \sqrt{2 m E_i}/\hbar$.  From this, we deduce
\begin{equation}
	{\rm var} ( M_{i \neq j} ) \simeq \frac{1}{A^2} \sum_{\bq\neq 0}
	\hV^2(\bq) \langle |W_{i \bq}|^2 \rangle  
	\langle |W_{j \bq}|^2 \rangle
\end{equation}

$|W_{j \bq}|^2$ can be interpreted as $(2\pi k_i)^{-2}$ times the
area of the intersection of two rings of diameter $k_i$ and width $2
\pi /L$, centered at a distance $|\bq| = q$.  Simple geometry
therefore gives, for $2\pi /L \leq |\bq| \leq 2k_i - 2\pi/L$ 
\begin{equation}
	\langle |W_{i \bq}|^2 \rangle \simeq \frac{4}{(q L)
		\sqrt{(2 k_i L)^2 - (q L)^2}} \;.
\end{equation}
We obtain for $i \simeq j$
\begin{equation}
     {\rm var} ( M_{i \neq j} )  \simeq 
        \frac{8}{\pi A^2} \int_{\pi/L}^{2k-\pi/L} \frac{dq }{q}
           \frac{\Vtf(q)^2}{(2kL)^2 - (qL)^2} \; .
\end{equation}

The variance of $M_{ii}$ and $N_{ij}$ and the covariance between
$M_{ij}$ and $N_{ij}$ can be computed along the same lines.


\bibliographystyle{hubprsty}
\bibliography{nanophys}

\begin{thebibliography}{10}

\bibitem{GrabDev92}
{\em Single Charge Tunneling: Coulomb Blockade Phenomena in Nanostructures},
  edited by H. Grabert and M. Devoret (Plenum, New York, 1992).

\bibitem{MesoTran97}
{\em Mesoscopic Electron Transport}, edited by L.~L. Sohn, L.~P. Kouwenhoven,
  and G. Sch{\"{o}}n (Kluwer, New York, 1997).

\bibitem{FerryGood97}
D.~K. Ferry and S.~M. Goodnick, {\em Transport in Nanostructures} (Cambridge
  University Press, New York, 1997).

\bibitem{Timp99}
{\em Nanotechnology}, edited by G. Timp (Springer-Verlag/AIP, New York, 1999).

\bibitem{KastnerRMP92}
M.~A. Kastner, Rev. Mod. Phys. {\bf 64},  849  (1992).

\bibitem{KouwenetalRev97}
L.~P. Kouwenhoven, C.~M. Marcus, P.~L. McEuen, S. Tarucha, R.~M. Westervelt,
  and N.~S. Wingreen,  in {\em Mesoscopic Electron Transport}, edited by L.~L.
  Sohn, L.~P. Kouwenhoven, and G. Sch{\"{o}}n (Kluwer, New York, 1997), pp.\
  105--214.

\bibitem{RalphRev97}
D.~C. Ralph, C.~T. Black, J.~M. Hergenrother, J.~G. Lu, and M. Tinkham,  in
  {\em Mesoscopic Electron Transport}, edited by L.~L. Sohn, L.~P. Kouwenhoven,
  and G. Sch{\"{o}}n (Kluwer, New York, 1997), pp.\ 447--467.

\bibitem{AleinBrouwGlaz01}
I.~L. Aleiner, P.~W. Brouwer, and L.~I. Glazman, in preparation (unpublished).

\bibitem{JalStoneAlh92}
R.~A. Jalabert, A.~D. Stone, and Y. Alhassid, Phys. Rev. Lett. {\bf 68},  3468
  (1992).

\bibitem{Stopa96}
M. Stopa, Phys. Rev. B {\bf 54},  13767  (1996).

\bibitem{Stopa98}
M. Stopa, Physica B {\bf 251},  228  (1998).

\bibitem{Hackenbroich97}
G. Hackenbroich, W.~D. Heiss, and H.~A. Weidenm\"uller, Phys. Rev. Lett. {\bf
  79},  127  (1997).

\bibitem{Alhassid98}
Y. Alhassid, M. G\"ok\c{c}edag, and A.~D. Stone, Phys. Rev. B {\bf 58},  7524
  (1998).

\bibitem{Vallejos98}
R.~O. Vallejos, C.~H. Lewenkopf, and E.~R. Mucciolo, Phys. Rev. Lett. {\bf 81},
   677  (1998).

\bibitem{Vallejos99}
R.~O. Vallejos, C.~H. Lewenkopf, and E.~R. Mucciolo, Phys. Rev. B {\bf 60},
  13682  (1999).

\bibitem{E2NBar99}
E.~E. Narimanov, N.~R. Cerruti, H.~U. Baranger, and S. Tomsovic, Phys. Rev.
  Lett. {\bf 83},  2640  (1999).

\bibitem{Kaplan00}
L. Kaplan, Phys. Rev. E {\bf 62},  3476  (2000), [arXiv:nlin.CD/0003013].

\bibitem{E2NBar01}
E.~E. Narimanov, H.~U. Baranger, N.~R. Cerruti, and S. Tomsovic, submitted to
  Phys. Rev. B, arXiv:cond-mat/0101034 (unpublished).

\bibitem{MarcusPatel98a}
S.~R. Patel, D.~R. Stewart, C.~M. Marcus, M. Gokcedag, Y. Alhassid, A.~D.
  Stone, C.~I. Duruos, and J. J.~S.~Harris, Phys. Rev. Lett. {\bf 81},  5900
  (1998).

\bibitem{ChangPeaks96}
A.~M. Chang, H.~U. Baranger, L.~N. Pfeiffer, K.~W. West, and T.~Y. Chang, Phys.
  Rev. Lett. {\bf 76},  1695  (1996).

\bibitem{MarcusFolk96}
J.~A. Folk, S.~R. Patel, S.~F. Godijn, A.~G. Huibers, S.~M. Cronenwett, and
  C.~M. Marcus, Phys. Rev. Lett. {\bf 76},  1699  (1996).

\bibitem{Mayer}
M.~G. Mayer, Phys. Rev. {\bf 60},  184  (1941).

\bibitem{Latter}
R. Latter, Phys. Rev. {\bf 99},  510  (1955).

\bibitem{Bohigas76}
O. Bohigas, X. Campi, H. Krivine, and J. Treiner, Phys. Lett. {\bf 64},  381
  (1976).

\bibitem{Sivan96}
U. Sivan, R. Berkovits, Y. Aloni, O. Prus, A. Auerbach, and G. Ben-Yoseph,
  Phys. Rev. Lett. {\bf 77},  1123  (1996).

\bibitem{Simmel97}
F. Simmel, T. Heinzel, and D.~A. Wharam, Europhys. Lett. {\bf 38},  123
  (1997).

\bibitem{MarcusPatel98}
S.~R. Patel, S.~M. Cronenwett, D.~R. Stewart, A.~G. Huibers, C.~M. Marcus,
  C.~I. Duruz, J.~S. Harris, K. Campman, and A.~C. Gossard, Phys. Rev. Lett.
  {\bf 80},  4522  (1998).

\bibitem{Simmel99}
F. Simmel, D. Abusch-Magder, D.~A. Wharam, M.~A. Kastner, and J.~P. Kotthaus,
  Phys. Rev. B {\bf 59},  R10441  (1999).

\bibitem{LuschEnsslin01}
S. L\"uscher, T. Heinzel, K. Ensslin, W. Wegscheider, and M. Bichler,
  arXiv:cond-mat/0002226 (unpublished).

\bibitem{BarOng01}
T.~T. Ong, H.~U. Baranger, C.~M. Marcus, and S.~R. Patel (unpublished).

\bibitem{Berkovits98}
R. Berkovits, Phys. Rev. Lett. {\bf 81},  2128  (1998).

\bibitem{Walker99a}
P. Walker, G. Montambaux, and Y. Gefen, Phys. Rev. Lett. {\bf 82},  5329
  (1999).

\bibitem{Walker99b}
P. Walker, G. Montambaux, and Y. Gefen, Phys. Rev. B {\bf 60},  2541  (1999).

\bibitem{Cohen99}
A. Cohen, K. Richter, and R. Berkovits, Phys. Rev. B {\bf 60},  2536  (1999).

\bibitem{Ahn99}
K.-H. Ahn, K. Richter, and I.-H. Lee, Phys. Rev. Lett. {\bf 83},  4144  (1999).

\bibitem{AshcroftMermin1}
N.~W. Ashcroft and N.~D. Mermin, {\em Solid State Physics} (Holt, Rinehart and
  Winston, New York, 1976), p.\ 344.

\bibitem{ReimannMann97}
M. Koskinen, M. Manninen, and S.~M. Reimann, Phys. Rev. Lett. {\bf 79},  1389
  (1997).

\bibitem{MartinLeb98}
I.-H. Lee, V. Rao, R.~M. Martin, and J.-P. Leburton, Phys. Rev. B {\bf 57},
  9035  (1998).

\bibitem{Wing99}
K. Hirose and N.~S. Wingreen, Phys. Rev. B {\bf 59},  4604  (1999).

\bibitem{Prus96}
O. Prus, A. Auerbach, Y. Aloni, U. Sivan, and R. Berkovits, Phys. Rev. B {\bf
  54},  R14289  (1996).

\bibitem{Berkovits97}
R. Berkovits and B. Altshuler, Phys. Rev. B {\bf 55},  5297  (1997).

\bibitem{BlantMirMuz97}
Y.~M. Blanter, A.~D. Mirlin, and B.~A. Muzykantskii, Phys. Rev. Lett. {\bf 78},
   2449  (1997).

\bibitem{BrouwOregHalp99}
P.~W. Brouwer, Y. Oreg, and B.~I. Halperin, Phys. Rev. B {\bf 60},  R13977
  (1999).

\bibitem{BarUllGlaz00}
H.~U. Baranger, D. Ullmo, and L.~I. Glazman, Phys. Rev. B {\bf 61},  R2425
  (2000).

\bibitem{MesoStoner00}
I.~L. Kurland, I.~L. Aleiner, and B.~L. Altshuler, Phys. Rev. B {\bf 62},
  14886  (2000).

\bibitem{JacquStPRL00}
P. Jacquod and A.~D. Stone, Phys. Rev. Lett. {\bf 84},  3938  (2000).

\bibitem{JacquStPRB01}
P. Jacquod and A.~D. Stone, cond-mat/0102029 (unpublished).

\bibitem{GutzBook}
M. Gutzwiller, {\em Chaos in Classical and Quantum Mechanics} (Spring-Verlag,
  New York, 1991).

\bibitem{AltSimHouches}
B.~L. Altshuler and B.~D. Simons,  in {\em Mesoscopic Quantum Physics}, edited
  by E. Akkermans, G. Montambaux, J.-L. Pichard, and J. Zinn-Justin (Elsevier
  Science, Amsterdam, 1995), pp.\ 1--98.

\bibitem{Strutinsky68}
V.~M. Strutinskii, Nucl. Phys {\bf A 122},  1  (1968).

\bibitem{Strutinsky72}
M. Brack, J. Damg{\o}ard, A.~S. Jensen, H.~C. Pauli, and V.~M. Strutinsky, Rev.
  Mod. Phys. {\bf 44},  320  (1972).

\bibitem{UllTatTomBar01}
D. Ullmo, T. Nagano, S. Tomsovic, and H.~U. Baranger, arXiv:cond-mat/0007330
  (unpublished).

\bibitem{QChaosHouches}
{\em Chaos and Quantum Physics}, edited by M.-J. Giannoni, A. Voros, and J.
  Zinn-Justin (North-Holland, New York, 1991).

\bibitem{BerryJ077}
M.~V. Berry, J. Phys. A {\bf 10},  2083  (1977).

\bibitem{Srednicki96}
M. Srednicki, Phys. Rev. E {\bf 54},  954  (1996).

\bibitem{Agametal97}
O. Agam, N.~S. Wingreen, B.~L. Altshuler, D.~C. Ralph, and M. Tinkham, Phys.
  Rev. Lett. {\bf 78},  1956  (1997).

\bibitem{AleinGlaz98}
I.~L. Aleiner and L.~I. Glazman, Phys. Rev. B {\bf 57},  9608  (1998).

\bibitem{Blant96}
Y.~M. Blanter, Phys. Rev. B {\bf 54},  12807  (1996).

\bibitem{BlantMir97}
Y.~M. Blanter and A.~D. Mirlin, Phys. Rev. B {\bf 55},  6514  (1997).

\bibitem{Leboeuf00}
P. Leboeuf and A. Monastra, Phys. Rev. B {\bf 62},  12617  (2000).

\bibitem{Wilkinson89}
M. Wilkinson, J. Phys. A {\bf 22},  2795  (1989).

\bibitem{WilkinsonAustin92}
E.~J. Austin and M. Wilkinson, Nonlinearity {\bf 5},  1137  (1992).

\bibitem{FyodMirl91}
Y.~V. Fyodorov and A.~D. Mirlin, Phys. Rev. Lett. {\bf 67},  2405  (1996).

\bibitem{Heinzel_private}
T. Heinzel, private communication (unpublished).

\bibitem{MarcusStewart97}
D.~R. Stewart, D. Sprinzak, C.~M. Marcus, C.~I. Duruoz, and J.~S. Harris,
  Science {\bf 278},  1784  (1997).

\end{thebibliography}

\end{multicols}

\end{document}